\documentclass{ws-ijmpd}
\usepackage[super,compress]{cite}
\begin{document}

\title{The Dark Sector Cosmology }

\author{Elcio Abdalla and Alessandro Marins}

\address{Instituto de Física, Departamento de F\'{\i}sica Geral, Universidade de S\~ao Paulo, \\
S\~ao Paulo, CEP 05508-090, Brazil,
eabdalla@usp.br\
bingotelescope@usp.br}

\maketitle

\begin{history}
\received{Day Month Year}
\revised{Day Month Year}
\end{history}

\begin{abstract}
 The most important problem in fundamental physics is the description of the contents of the Universe. Today, we know that 95\% thereof is totally unknown. Two thirds of that amount is the mysterious Dark Energy described in an interesting and important review\cite{CopeSamiTsu}. We briefly extend here the ideas contained in that review including the more general Dark Sector, that is, Dark Matter and Dark Energy, eventually composing a new physical Sector. Understanding the Dark Sector with precision is paramount for us to be able to understand all the other cosmological parameters comprehensively as modifications of the modelling could lead to potential biases of inferred parameters of the model, such as measurements of the Hubble constant and distance indicators such as the Baryon Acoustic Oscillations. We discuss several modern methods of observation that can disentangle the different possible descriptions of the Dark Sector. The possible application of some theoretical developments are also included in this paper as well as a more thorough evaluation of new observational techniques at lower frequencies and gravitational waves.
\end{abstract}


\ccode{PACS numbers:}


\section{The Standard Cosmological Model and the Dark Sector}	

The description of the Universe has always been a major and deep problem in the History of Human Understanding, as well as in the framework of Modern Physics based on Einstein Gravity. The cosmological solution, e. g. the Friedmann-Lemaitre-Robertson-Walker (FLRW) solution \cite{Dodelson} was the first attempt to describe the Universe as a whole as a consequence of a physically well established theory, that is, Einstein gravity.

However, since 1933 a new ingredient was little by little introduced in the proposed theoretical structure of the cosmos: a previously unknown new element in the cosmos, in the beginning thought to be as largely massive as at least 10 times the usual matter density.

For a long time the question of whether there exists a hidden Dark Sector  was a burden for a high level understanding of the Universe, since no natural representative from Elementary Particle Standard Model \cite{StandardModel} has ever represented such a new Cosmological sector. As a matter of fact, this is, in most aspects, a still standing problem.

The seminal paper about a Dark Sector \cite{Zwicky} (confirmed by others at the time \cite{Oort,Ostriker})  was based on the movement of galaxies in the Coma cluster (Zwicky actually named Dark Matter at the paper), while in recent times several different observations point in the same direction \cite{Dodelson}. As we know it today, in the so called concordance model, the Dark Sector is constituted by two parts, Dark Matter (DM) and Dark Energy (DE), the former having the usual meaning of gravitationally interacting mass, but oblivious to other standard Particle Physics interactions, while the latter is a new strange kind of object leading to the accelerated expansion of the Universe, as predicted by observations, by means of a (very) negative pressure, $\rho+3p/c^2\le 0$.

The choice $p=-\rho c^2$ corresponds to the addition of a cosmological constant $\Lambda g_{\mu\nu}$ to the right hand side of Einstein Equations, that is\cite{Dodelson,Weinberg},
\begin{equation}
    R_{\mu\nu}-\frac 12 g_ {\mu\nu}R = 8\pi G T_{\mu\nu} -\Lambda g_{\mu\nu}\quad ,
\end{equation}
corresponding to the action
\begin{equation}
    S=\frac 1{16\pi G}\int R\sqrt {-g} +\frac{\Lambda}{8\pi G}\int \sqrt{-g} +S_{matter}\quad .
\end{equation}
The Standard Cosmological model is based on the above action for $S_{matter}$ corresponding to the Elementary Particle Standard Model and some heavy massive fluid describing DM. Moreover, the Cosmological constant describes the acceleration of the Universe, a fact corroborated by observations. The cosmological constant has to be, in the above formulation, a Universal constant in view of covariance.

There are two basic theoretical problems in the Cosmological Standard Model. The first is the fact that $\Lambda$ is too small as compared to any Field Theory attempt to describe it in terms of Standard Field Theory \cite{WeinbergRMP}. The second is that both DM and DE are equally important at the present era in spite on their different behavior as predicted by the Standard Cosmological Model. This is the coincidence problem \cite{pavon}.

Therefore, it is natural to consider models of DM and DE with non trivial characteristics, that is, not just a fluid and a constant. Actually, this is a fundamental question that should lead to a formulation of a more fundamental theory of these elements, possibly increasing the scope of the Standard Model of the Elementary Particles.

\section{The Generalized Dark Sector}
Several alternative descriptions of DE by itself are very well described in the classical review by Copeland, Sami and Tsujikawa \cite{CopeSamiTsu}. There are, in fact, several remarkable puzzles about the very existence of DE. Various descriptions and models have been proposed to describe an accelerating Universe in the above quite remarkable review. Nonetheless, no specific model has, up to now, prevailed as a unified and final (or preferred) description of DE.  In that review, the huge number of alternatives for the explanation and structure of Dark Energy was  presented. The case of a Cosmological constant is observationally the simplest, but to cope with its small value is very difficult and one might refer even to an Anthropic principle (see section IV of the review for details). Thus, there are several models of Dark Energy, by far the simplest and most popular is via scalar fields. On the other hand, modifications of gravity are also very popular and have been discussed in full detail in some parts of the review. String theory has the ability to propose some hints, but now, in this direction, there are very constrained possibilities. Nonetheless, a few string inspired models such as DGP may be considered \cite{Yinetal}. A Dark Sector as a consequence of a hidden string sector is also possible \cite{Honecker,Acharya2018}.

After models describing DE from a phenomenological holographic point of view \cite{wanggongabdalla}, the simplest description of the Dark Sector is that of an unknown Dark Matter heavy fluid and a Cosmological constant, what is known as the Standard Cosmological Model, or $\Lambda$CDM model, which, as above mentioned, has problems in its theoretical interpretation. However, if we leave the realm of a non interacting fluid describing DM and a (Cosmological) constant, the most conservative possibility is the description of these elements by means of a field theory, unavoidably leading to a variety of models of an interacting Dark Sector, that is, a set of fields in the Hidden Sector, or the even more complicated possibility of a modification of Einstein gravity itself.

These possibilities are very appealing from the more theoretical point of view, and lead to the idea of an independent Dark Sector, very weakly coupled to the baryonic sector, or not coupled  at all except for gravity, but with an independent dynamics \cite{WangAbdAtPav}.

From the point of view of string theory and beyond, a Dark Sector can be naturally accommodated. As a matter of fact, we can say that a Hidden Sector is basically mandatory in that context. One obvious case concerns the $E_8\times E_8$ heterotic string, but a Hidden Sector arising from several different compactification schemes have been largely discussed in the string literature \cite{Honecker,Acharya2018}.

Usually, the preferred type of DM consists in taking the Lighest Supersymmetric Particle in the Visible Sector (LSVP) \cite{Hooper}. The advantage consists in the fact that being the lightest it must be stable and interacts weakly with visible matter. There are several candidates, the preferred one being a Neutralino. However, there are several other candidates, highly discussed in the literature \cite{GBertone}. Nonetheless, it has been found recently that it might be unstable \cite{Acharya2016}.

Taking for granted the existence of an interaction in the Dark (Hidden) Sector, it will be, by the same token, natural to consider a Dark Energy arising from another Hidden Sector field. But what field? A natural possibility allowing for a negative equation of state is a scalar field, where the potential quite naturally leads to a negative contribution to the pressure. In the string literature scalars are abundant.

Therefore, it is natural to consider models of the Dark Sector consistent with the idea of describing Dark Matter by a fermion and a scalar describing Dark Energy \cite{CostaOlivariAbdalla}. Possibilities of introducing vector fields, as implied by some string inspired models is also feasible \cite{KoivistoMota}. Moreover, fermions can easily accommodate the idea of Cold Dark Matter.

\section{State of the Art Observational Methods in Cosmology}
From the point of view of observations, constraints on Cosmology are, nowadays, much stronger. We are in a so called precision cosmology age. Data arise from different directions:

1. {\it Dark Energy Direct Observation: Structure of the Hubble law determining  cosmological acceleration for relatively small Redshift data ($z<1$)}
   
   Supernovae observations are well suited, since some time, to confirm and improve Hubble's result. It has been shown from Supernovae observations \cite{SuperNovae} that to more than $7\sigma$ the data are compatible with an accelerating Universe as described by a positive Cosmological Constant (CC). The amount of DE (CC) is of the order of 2/3 of the critical value. The results are, observationally, very well grounded and obtained by different groups.
   
2. {\it Dark Matter amount from direct observation.}
   
   Direct Dark Matter observations have been done since almost a century and continue to be performed. Although it has led people to propose alternatives such as MOND (modified gravity), the existence of a Dark Sector is, from the side of DM, basically undeniable. Indeed, modifications of gravity at large scales can explain higher speeds of astrophysical objects around a galaxy or a cluster. However, Modified Gravity cannot explain the CMB anisotropies distributions, which require a definite amount of Dark Matter. Constraints from the Microwave Background observations are very strong, especially now after two decades of detailed observations that we comment below.

 3. {\it Constraints on cosmological parameters from CMB observations (COBE, WMAP and, more recently, Planck)}
   
   The observed power spectrum from Cosmological Microwave Background (CMB) is one of the finest observations in nature. CMB is the most perfect Black Body, characterized by a temperature of T$\approx 2.726$K. Fluctuations of 1 part in 100.000 are predicted to perfection from cosmological perturbations, which depend on the cosmological parameters, mainly Matter, Dark Matter and Dark Energy relative amounts (besides eventual curvature, which is finally consistent with zero as a result of such observations). 
   
   The CMB prediction is rooted on the two point function of the density fluctuations (power spectrum) depending heavily on all cosmological parameters. The results for the power spectrum as a function of the angular parameter in Fourier space is well know and can be compared with observations, providing tight constraints on the cosmological parameters. As a result, the latest Planck Observations imply $\Omega _{DE} = 0.685$ and $\Omega_{M} = 0.315$, with total matter consisting of $85\%$ of DM and $15\%$ baryonic matter, for the $\Lambda \textrm{CDM}$ model. When curvature is allowed to vary its best value is $\Omega_{K} = -0.011$, compatible with 0 by one $\sigma$\cite{Planck2018_parameters}. This is also consistent with the inflationary scenario.
   
   4. {\it Constraints from Baryonic Acoustic Oscillations as standard rules.}
   
   Some years ago, it has been established that relic waves left by the baryonic electric interaction previous to decoupling could now be observed as giant correlations of the rough size of 150 Mpc \cite{Eisenstein}, the so called Baryonic Acoustic Oscillations (BAO). The size of that wave stands as a standard ruler allowing for a new set of constraints on cosmological parameters. 
   
   5. {\it Observations from weak lensing effects.}
   
   Lensing as an effect of General Relativity is one hundred years old and reminds us of the first confirmation of General Relativity. Now, the consequence of the well known lensing process \cite{Weinberg} arising from intervening matter with respect to observation has an important effect in the power spectrum \cite{Lensing}.

  6. {\it Results arising from Redshift Space Distortion.}
   
   The most precise measurement we can have in cosmology is the redshift. It is directly related to distance as a consequence of Hubble law \cite{SuperNovae}. However, local movements can interfere with the exactness of the relation, leading to distortions in view of local relative motions as e.g. inside a galaxy, what leads to distortions in the redshift map \cite{RSD01}. This is an important piece of information to deal with large scale structure maps.
   
  The redshift-space distortion
(RSD) has been measured in the past few years with higher precision \cite{RSD02}. It can give information about the large scale structure and cosmological parameters of relevance for model building \cite{RSD03}, because the local movement of galaxies towards higher density distribution implies a peculiar component to the redshift as compared to the Hubble predicted value. It is thus a  powerful complementary observation to break the possible degeneracy in cosmological models. This is also because the dynamical growth history in the cosmological structure can be distinct even in case they display the same  evolution in the background.  It is expected that including the large scale structure information by adding the RSD measurements can provide a rich background on theoretical models of cosmology \cite{RSD03}.

 7. {\it Optical data constraining Large Scale distribution of Matter.}
    
    The above point is related to the structure of perturbations at large scales. Perturbation of Einstein Equations can have integer spins from zero to two. Spin 2 perturbations concern gravitational waves. Vector perturbations decay quickly and are not cosmologically important. Scalar perturbations remain as the template for density fluctuations.
    
    Today it is common to have maps of the Universe from observations. From the comparison of such so called mocks with the gravitational theory supplemented by the matter model used can provide a way to constrain the cosmological parameters.

8. {\it Future Developments }

In the future, constraints from gravitational waves observations can be incorporate in this list. Indeed, we can show that observations of these waves can be used as standard sirens to improve Hubble data (see below).

By the same token, also in the future, neutrino observations, especially neutrino masses, can be used as constraints in Cosmology \cite{filipe}.
   
9. {\it Observations of the
21 cm line by means of the technique of Intensity Mapping (what overlaps with some of the previous results)}

Hydrogen is the most common visible element (baryonic) of the Universe. A map of the Hydrogen distribution is an valuable template of matter and much easier to obtain as we shall see. It is also natural to suppose that the Hydrogen distribution may also be a good template of Dark Matter, assuming that matter in general follows a standard (possibly biased) distribution. 

Usually, the Large Scale Structure by means of a redshift survey is an expensive method, and alternatives are searched, see for instance the JPAS project and its role in the Dark Sector search \cite{jpas}.  

On the other hand, the Hydrogen atom emits a very typical radiation corresponding to the hyperfine transition, the 21 cm line. Although the higher state being quite stable, Hydrogen is common enough to provide a way to measure distribution by the technique of Intensity Mapping (IM) \cite{IM}, that is, by the average intensity of the 21cm line. 

The 21cm emission of the underlying matter, can be 50 times more efficient than the previous technique using individual galaxies \cite{AbdallaRawlings}. Therefore, progress  can  be  made  with  much smaller scale instruments, as for instance BINGO \cite{BINGOprojectpaper}. Today, several projects in Radio Astronomy aim using this method to probe the Dark Sector \cite{WangAbdAtPav}.

\section{Statistical Methods in Cosmology}
Cosmology is the result of solutions of the Einstein Field Equations coupled to the Standard Particle Model. Therefore, it can be seen as a classical solution of a highly nonlinear field theory, including the particle fields. The metric and possibly phenomenological fluids are usually used in the impossibility of directly describing certain sophisticated solutions from first principles in a general field theory.

Given the initial conditions from inflation and its probabilistic character, the density profile and correlation functions are the main objects we can aim to observe and describe. A sophisticated and detailed cosmology is described by correlators, whatever mathematical space is used to describe it (real, Fourier, harmonic, wavelets, among others). Hard data analysis  is necessary to extract as accurate as possible cosmological information.

Cosmological surveys, in current and next generations,  explore a huge amount of data, with increasingly larger areas and greater analysis depth as well as greater resolution. Sophisticated data analysis methods are needed to deal with the volume and accuracy of the information. As an example, the next generation of galaxy surveys is expected to obtain up to billions of galaxies\cite{Abdalla2015}.

\section{Intensity Mapping: Exploring 21cm line cosmological information}
The standard approach to probing Large Scale Structure (LSS) by means of a large galaxy-redshift survey has been extensively used in the optical and far infrared bands.  Galaxies are  used as tracers of the underlying total matter distribution. However, galaxy surveys have a threshold for a minimum flux\cite{M.Santos2015}. Only galaxies above this value can be individually detected. These surveys require a large integration time of observation to obtain good determination of the galaxy redshifts from their optical spectra\cite{D.Alonso2014}. Considering the epoch before the existence of structure, such as reionisation or even before, it is necessary to appeal to other types of survey.

In the past two decades another method has been developed and used to explore the radio band, called Intensity Mapping (IM), that explores a larger area of the sky in a shorter time. Instead of galaxy surveys, IM does not require identifying individual objects thus not having a threshold of the minimum flux\cite{Wyithe2009}. IM, instead of identifying galaxies, measures the total flux from many galaxies which will underline matter. Therefore, we can be use HI as a matter tracer. Observing 21 cm HI emission we can explore a long time range of the life of the Universe since ionization epoch. In post-reionisation (low redshift), most of (ionised) hydrogen are outside galaxies and HI are inside galaxies in  high-density gas clouds systems that are shielded from ionised UV photons\cite{Villaescusa-Navarro2018,M.Santos2015}.

IM can be performed by low-resolutions interferometers or by single-dish experiments. There are several examples, CHIME \cite{A.Liu2019} and SKA\cite{Abdalla2015, D.Bacon2018}  are interferometers. In the case of SKA (in development) that will cover a redshift range until z $\sim$ 6; that is, until reionisation epoch. For "dishes" experiments there are MeerKAT\footnote{http://meerkat2016.ska.ac.za/}  and BINGO\cite{BINGOprojectpaper} that will cover different sky regions and redshift ranges post-reionisation.

In the reionisation epoch the 21cm power spectrum is highly dependent on the details of the reionisation process and post-reionisation although it does not depend of the description of the process, the signal is very weak, since most of the HI is destroyed during the reionisation epoch\cite{A.Hall2013}, and the signal is highly contaminated. Therefore, the main challenge for HI IM surveys are the astrophysical contamination and the systematic effects about the observed signal. To reconstruct the 21cm signal techniques already used in CMB experiments classified as \emph{Component Separation} are necessary. The 21cm IM has the advantage that the astrophysical components vary smoothly with frequency and they have stronger signal than the 21cm line, facts that can be used to reconstruct the signal. Different techniques exploring different characteristics about foregrounds or about 21cm signal are available or under development to disentangle the signals. As an example we cite GNILC\cite{Olivari2018_1, Olivari2018_2, Karin2020}  and GMCA\cite{I.Carucci2020} that are also used in case of the CMB\cite{Planck2018_foregrounds}.

Other lines and atoms or molecules can also be used in IM such as CO(1-0)\footnote{https://kipac.stanford.edu/research/projects/co-mapping-array-pathfinder-comap}, CO(2-1), CO(3-2) among others \cite{J.Fonseca2019}.

\section{Dark Energy and BAO from 21cm IM}

The accelerated expansion of the Universe is a formidable challenge. What mechanism or component is actually causing this behavior is a big puzzle and an extraordinary way to study it is through BAO at low redshifts, when, according to what studies and analyses of the data have shown us, the DE becomes relevant in the dynamics of the expansion of the Universe. The post-reionisation epoch is a valuable source of information about the nature of DE and can be studied from BAO, that is, an information from a primordial epoch of the Universe. 

There was a period when baryons and photons were coupled, building a hot plasma. The Electromagnetic coupling leading to Compton scattering builds patterns in the distribution of the matter and photons, but with the expansion of the Universe and its cooling the patterns decrease and freeze when photons and baryons decouple. These patterns can be seen both in the CMB spectrum and in the galaxy spectrum today. Considering a spherical perturbation of matter density, it will propagate outwards (in relation to the center of the perturbation) with a \emph{sound speed} proportional to the light speed. It is redshift dependent. That is, the perturbations spread as sound waves on this plasma and stops its propagation when baryons and photons decouple. Such a process leads to a characteristic peak in the matter correlator that, when seen in Fourier (or harmonic) space, appears as damped oscillations in a given region of scale. The phenomenon is called Baryon Acoustic Oscillation (BAO). In real space that peak distance is about 150 Mpc and is called the sound horizon corresponding to the distance travelled by the sound wave until shortly after decoupling (drag epoch). HI, being a matter tracer, must display such a pattern too and this property must appear in its power spectrum. In this scenario, the BINGO telescope \cite{BINGOprojectpaper} aims to detect BAO in radio frequency. This is how the 21cm IM method can be used.

But how exactly can we use BAO to study Dark Energy? BAO carried information about the primordial Universe and in the corresponding redshift of the measurement. In general, cosmological length, and therefore distance, depends on what the Universe is made of and in what proportions. The sound horizon length depends on the history of the Universe until the drag epoch and its distortions measured in radial and transverse directions also depend on the properties of the Universe. Still, sound horizon is a fixed value of the comoving length for all times after drag epoch and can be used to calibrate distance, that is, it is a \emph{standard cosmological ruler}\cite{C.Blake2003, Bassett2010}. Thus, BAO can be used for both astronomical calibration and cosmological parameters constraints.

In the BINGO case, the angular power  spectrum  will be used as data and BAO appears as fluctuations in a given multipole range. However, in order to see this fluctuation, the component separation must be as accurate as possible, such that contamination and errors in the data do not suppress fluctuations. In addition to foregrounds contamination, that is about a thousand times stronger than 21cm signal, there are contaminations such as polarization leakage\cite{I.Carucci2020}, 1/f noise\cite{S.Harper2020}, temperature system, Radio Frequency Interference (RFI), among others, that are smaller than foregrounds, but also very important contaminations that need to be precisely understood for the pipeline construction \cite{Karin2020} and must be properly extracted. 

The large covered area and modern design make BINGO, in addition to a great cost benefit, a rich source of new information, important in itself and also important to be treated together with the data already available, with a coverage of 5400 deg$^2$ and in the range z = 0.13-0.48, exploring the galactic south through two dishes of approximately 40 m in diameter and the signal collecting horns\cite{Abdalla2020}.

\section{New Trends in Cosmology and Astrophysics}

\subsection{Gravitational Waves}
For about 100 years, gravitational waves eluded physicists: it had not been possible to obtain a connection between theory and experiments, the latter were lacking useful results. The only hint to gravitational wave was the loss of energy in a binary system.

In the last few years enormous advances have been made in the quest of gravitational waves, which have been finally explicitly encountered. But this is not the only good news, it can also be used in cosmology: the amplitude of gravitational waves is related to the measure of the luminosity distance from the source. This implies a possibility to establish Hubble law in an independent way, that is, obtaining a new redshift/distance diagram. We say that  gravitational waves can be used as standard sirens, as in gravitational waves originating from the collision of a pair of neutron stars. A few years ago, (GW170817) and its electromagnetic counterpart (GRB170817A) were both detected. 

Therefore, Gravitational Waves can also be used to investigate cosmology (and consequently also the Dark Sector) besides electromagnetic based observations.

The   luminosity distance in a  flat space-time is given by the expression
\begin{equation}
\label{dLz}d_L(z)=\frac{c(1+z)}{H_0}\int_0^z\frac{dz'}{E(z', \vec{\Omega})}\quad,
\end{equation}
where $E(z,\vec{\Omega})=H(z,\vec{\Omega})/H_0$ is the normalized Hubble function, depending on the redshift $z$ and the parameter set $\vec{\Omega}$ characterizing the cosmological model (relative densities). 

Gravitational Wave amplitudes for a binary system depend on the chirp mass, $\mathcal{M}_c\equiv M\eta^{3/5}\equiv (m_1+m_2) \eta ^{3/5}$ where $\eta=m_1m_2/M^2$. It can be measured by the Gravitational Wave signal phasing \cite{Zhao2010, Cai2016}, defining  the luminosity distance  $d_L$ from the GW amplitude. The relation between the luminosity distance and the amplitude is the core of the importance of gravitational waves for cosmological investigation, we need just one of them, and it determines the other object. 

Interferometers can measure the $h(t)$. In a gauge described by ``plus"  and ``times" modes $h_+$, $h_{\times}$, the strain is given by
\begin{equation}
\label{strain}h(t)=F_+(\theta,\phi,\psi)h_+(t)+F_{\times}(\theta,\phi,\psi)h_{\times}(t)\; ,
\end{equation}
where the $F_{+,\times}$ characterize the beam, $\psi$ is the polarization angle and $(\theta, \phi)$ give the position of the source.

The observed chirp mass is related to the physical chirp mass by a redshift factor,  $\mathcal{M}_{c,obs}= (1+z)\mathcal{M}_{c,phys}$. The Fourier transform $\mathcal{H}(f)$ of the strain $h(t)$ is
\begin{equation}
\label{Hfourier}\mathcal{H}(f)=\mathcal{A}f^{-7/6}e^{i\Psi(f)}\quad,
\end{equation}
where $\Psi(f)$ is a phase and the amplitude is proportional to a power of the chirp mass, $\mathcal{A} \sim \mathcal{M}_c^{5/6} $.

We can generate a mock catalog $d_L-z$ by coalescence of BNS pair in the mass range $[1-2]M_{\odot}$ for each individual neutron star. The redshift distribution of the observable sources follow a calculable function \cite{Belgacem2019} in terms of another function describing the redshift evolution of the burst rate. We have to use also  the formation rate of massive binaries and the delay time distribution $P(t_d)$ \cite{Belgacem2019} which describes the minimal delay time for a binary system to evolve to merger. The cosmic star formation rate based in Gamma-Ray Bursts (GRB) rate \cite{Vangioni2014} is also needed. We can adopt $R_{{BNS}}(z=0)= 920~{Gpc}^{-3} {yr}^{-1}$ as estimated by  LIGO/Virgo observation run with the assumption that the mass distribution of neutron stars follows a gaussian mass distribution.

The Hubble expansion can be estimated in an independent way and a new cosmological test can be obtained. Therefore, gravitational waves as standard sirens can be a very useful cosmological probe in the near future. Third generation detectors as the case of the Einstein Telescope can improve current Gravitational Waves observations and have sensibility to detect an order of $10^2$ events per year, which is enough to impose constraints as good as the current cosmological probes.

Recently, forecasts about interacting Dark Sector cosmology has been proposed with good results pointing to further constraints. As a result there is a decrease of almost $\sim 90\%$ in error in relation to CMB data only, which suggests that standard sirens can help solving the tension in $H_0$ between CMB and Supernovae in the near future.  Some of the results have been obtained by Yang et al and Rhavia et al  \cite{Yang2019,Rhavia2020} for interacting vacuum-energy models,  as well as in interacting DE/DM models \cite{DEDMInteraction}. Authors \cite{Yang2019} found an improvement of $17\%$ in the coupling  and $35\%$ in $H_0$ with the addition of GW simulated data to CMB+BAO+SN data, and the reduction of the uncertainty on the DM-DE coupling by a factor of 5 \cite{Yang2019_2}. 

This shows a very intersting interplay of different types of physics pointing directions in physics and cosmology understanding.

\subsection{Other Astrophysical Objects}

New physics is at the verge of being discovered. Even clearer is the fact that new objects in the sky generally provide new constraints in our description of the cosmos. 

Until some time ago, the only information about the sky came from the direct optical observation. Now, with further knowledge and observations of neutrino properties, gravitational waves as above, Gamma Ray Bursts and other minor objects, further data for cosmology reconstruction shall follow.



\subsection{Fast Radio Bursts}

Discovered in 2007 \cite{Lorimer2007}, FRBs are short bursts ($0.04-161$ Jy) of radio photons  of unknown  origin \cite{Pen2018,Platts2018} and several possible explanations, a question left undecided up to now. The  time delay of photons with different  frequencies is characterized by the so called Dispersion Measure, $\int n_e dl $, which basically informs us how much space it has travelled. From the dispersion measure we know that FRBs are extra galactic objects. Today more than a hundred such events have been observed; several are knew. If we are able to define their position we are going to be able to have a further test of cosmological parameters. Direct observation from Radio Telescopes (BINGO is one such, there are several others \cite{BINGOprojectpaper,SKA}) is going to be possible. In case we know their position with accuracy (some outriggers should be enough adjoining a Radio Telescope as BINGO) we shall be able to know their host galaxy, therefore their distance, allowing us to characterize much of their parameters.

Several FRBs with frequencies from 300 MHz to 8 GHz have been recently observed \cite{Gajjar2018,Chawla2020}. We can use this type of information not only to constrain cosmological parameters  \cite{Zhou:2014yta} but also to test the Weak Equivalence Principle \cite{Wei:2015hwd} as well as getting constraints on the photon mass \cite{Wu2016} and  compact DM \cite{Wang:2018ydd}. We can consider  the Inter Galactic Medium \cite{Deng:2013aga} and foresee other possible applications \cite{LinderLandim}.

FRBs have been seen in various experiments such as the Australian Square Kilometre Array Pathfinder  \cite{Bannister:2017sie}, the  Parkes Radio Telescope   \cite{Lorimer2007},  the CHIME \cite{Amiri:2018qsq}, the UTMOST telescope \cite{Caleb:2017vbk}, the GBT, Arecibo and Apertif  \cite{Masui:2015kmb,Spitler:2014fla,Connor:2020oay}.

Radio Telescopes at the verge of starting to work will be able to make several observations of FRBs. In particular, the BINGO telescope will possibly be able to evaluate its host galaxy, thus important cosmological conclusions can be drawn from them, especially concerning the structure of acceleration and thus Dark Energy, besides the formidable problem of the astrophysical structure of FRBs.

\section{Conclusions}

Today, cosmology is possibly the hottest area of research in fundamental physics. It can provide us a view of a world at least 20 times larger than ours (with respect to its content), in case the Dark Sector has the same richness as the Baryonic Sector, what is not beyond all possibilities. In this direction, DE is the most tantalizing information, certainly beyond our common sense knowledge. In particular, DE denies the Strong Energy Condition, a property quite unthinkable in the near past!

In view of these properties, it is no doubt that the physical properties of a purported Dark Sector are (or can be) very peculiar and encompasses non trivial modifications of Standard Particle Theory and models. Previous important results seem to pale before these properties.


\section*{Acknowledgments} We would like to thank FAPESP and CNPQ (Brazil) for long standing financial support.


\end{document}